# Derivation of a non-stoichiometric 1/1 quasicrystal approximant from a stoichiometric 2/1 quasicrystal approximant and maximization of magnetocaloric effect

Farid Labib[1*], Hiroyuki Takakura[2], Asuka Ishikawa[1], Takenori Fujii[3] and Ryuji Tamura[4*]

[1] *Research Institute of Science and Technology, Tokyo University of Science, Tokyo 125-8585, Japan*

[2] *Division of Applied Physics, Faculty of Engineering, Hokkaido University, Sapporo 060-8628, Japan*

[3] *Cryogenic Research Center, The University of Tokyo, Bunkyo, Tokyo 113-0032, Japan 980-8577 Japan*

[4] *Department of Materials Science and Technology, Tokyo University of Science, Tokyo 125-8585, Japan*

(Date: June 5, 2025)

The present research introduces a novel strategy for tuning magnetic properties by overcoming the compositional limitation of stoichiometric intermetallic compounds via extension of their stability into a new dimension within valence electron-per-atom ($e/a$) parameter space. Focusing on approximant crystals (ACs), a "double hetero-valent elemental substitution" is employed in a stoichiometric Ga–Pt–Gd 2/1 AC whereby $e/a$ is lowered from 1.92 to 1.60. Through this approach a new family of stable Ga-based Tsai-type 1/1 ACs with exceptionally wide composition stability within $e/a$ space is derived. Remarkably, magnetic ground state is altered from initially spin-glass to ferromagnetic (FM) with second order phase transition and mean-field-like critical behavior. More importantly, through this strategy, the isothermal magnetic entropy change ($\Delta S_M$) enhanced significantly and reached a maximum value of $-8.7$ J/ K mol-Gd under a 5 T magnetic field change, even comparable to leading rare-earth magnetocaloric materials including $RCo_2$ phases. These findings demonstrate the high potential of a double hetero-valent elemental substitution for tailoring magnetic properties and magnetocaloric response in stoichiometric compounds, offering a new pathway for designing high-performance magnetic refrigeration materials even beyond the quasicrystals and ACs.

## I. INTRODUCTION

The number of valence electrons per atom ($e/a$) has long been considered as a critical parameter governing the electronic structure and stability of intermetallic compounds, as certain structures such as body centred cubic, complex cubic lattices including γ-Brass phases, hexagonal close-packed phases and quasicrystals (QCs) are known to be stabilized at particular $e/a$ values ($\approx 1.5, \approx 1.62, \approx 1.75, \approx 2.00$, respectively) [1,2]. Historically, the influence of $e/a$ on magnetic properties has typically been discussed through the Slater–Pauling principle [3], which relates the magnetic moment to the number of valence electrons.

In recent years, such $e/a$-based design principle has played a crucial role in controlling the magnetic properties of certain class of complex alloys. A notable example includes the *non-stoichiometric* Au–SM–R (SM: semimetal, R: rare-earth) Tsai-type 1/1 approximant crystals (ACs), wherein the Au/SM ratio can be varied over a wide range of ∼ 33 at.% without affecting the underlying crystallographic symmetry (see for example [4]). Reducing $e/a$ ratio in these 1/1 ACs has been shown to induce long-range ferromagnetic (FM) [5–8] and antiferromagnetic (AFM) [9–12] orders with intriguing non-coplanar whirling spin structures [13,14]. More recently, the emergence of FM [15,16] and AFM [17] orders have also been evidenced in real p-type icosahedral QCs.

While remarkable achievements have been made in controlling magnetic behavior via $e/a$-tuning as discussed above, one of the modern challenges in the study of intermetallic compounds, including but not limited to QCs and ACs, remains unaddressed. The challenge is that the inherent stoichiometry of many alloys severely limits degrees of freedom in compositional, and consequently $e/a$ tuning which is essential for tailoring magnetic properties. Many known QCs and ACs, for instance, have an $e/a \approx 2.00$ [2] and often exhibit spin-glass behavior as a result of geometric frustration and competing interactions at this particular $e/a$ value, as in Zn-Mg-$R$ [18–20] and Cd-Mg-$R$ [21–25] iQCs/ACs.

In the present work, a novel materials design strategy is introduced to overcome this limitation by extending the stability of stoichiometric compounds into a new dimension within $e/a$ space, thereby enabling the magnetic properties tuning. Following this approach, we report the development of entirely new family of FM Tsai-type Ga-Pt-Au-Gd 1/1 ACs from a parent spin-glass stoichiometric $Ga_{52}Pt_{34}Gd_{14}$ 2/1 AC through "*double hetero-valent elemental substitution*" leading to drop of $e/a$ from 1.92 to 1.60.

The resulting quaternary 1/1 ACs are stable and exhibit second order FM transition with mean-field-like critical behavior around Curie temperature ($T_C$) across a wide $e/a$ space. The isothermal magnetic entropy change ($\Delta S_M$) reached a maximum value of $-8.7$ J/ K mol-Gd under an applied field change of 5 T in the extended $e/a$ space, indicating the approach adopted in the present work as a robust and effective strategy for tuning magnetic properties and $\Delta S_M$ in stoichiometric compounds in a way that was previously impossible. The present work opens new pathways for the design of high-performance magnetic materials such as magnetic refrigeration materials.

…

## II. EXPERIMENT

The synthesis protocol includes arc-melting of the constituent elements under the argon atmosphere. Each sample was melted several times to ensure homogeneity. Following arc-melting, the samples were annealed at 1073 K for 50 hours in an argon-filled quartz tube to homogenize the microstructure. The phase purity of the samples after synthesis was confirmed through powder x-ray diffraction (XRD) using Rigaku SmartLab SE X-ray diffractometer with Cu-Kα radiation. A representative sample with a nominal composition of $Ga_{33}Au_{33}Pt_{20}Gd_{14}$ was selected for room temperature single crystal x-ray diffraction (SCXRD) experiment with the aim of crystal structure analysis using an XtaLAB Synergy-R single-crystal diffractometer equipped with Hybrid Pixel Array Detector (HyPix6000, Rigaku) with Mo Kα radiation ($\lambda = 0.71073$ Å). The SCXRD data were refined assuming $Im\bar{3}$ as the space group, utilizing SHELXT [26] for initial model generation and SHELXL [27] for the final refinement. Crystallographic data and refinement parameters as well as the details of the final model including atomic coordinates, Wyckoff positions, site occupations, and equivalent isotropic displacement parameters ($U_{eq}$) are listed in Table S1 and S2 of the Supplemental Material.

For bulk magnetization measurement, a superconducting quantum interference device (SQUID) magnetometer (Quantum Design, MPMS3) was utilized under zero-field-cooled (ZFC) and field-cooled (FC) modes within a temperature range of 1.8 K to 300 K by applying external dc fields up to $7\times10^4$ Oe. Additionally, specific heat measurements were conducted in a temperature range of 2 K – 40 K by a thermal relaxation method using a Quantum Design Physical Property Measurement System (PPMS). The $\Delta S_M$ of the samples are derived from isothermal magnetization measurements conducted up to $\mu_0 H = 7$ T at various temperatures using the thermodynamic Maxwell relation, as will be discussed later.

## III. RESULTS

In this work, the stoichiometric $Ga_{52}Pt_{34}Gd_{14}$ 2/1 AC is selected as a parent compound for a double hetero-valent elemental substitution, whereby Ga and Pt are partially exchanged with Au. Such substitution not only lowers the order of AC from 2/1 to 1/1 but also fundamentally breaks the stoichiometric constraint and expands the compositional degrees of freedom. Consequently, the 1/1 AC becomes non-stoichiometric, which allows its stability to span a broader valence $e/a$ range. In other words, the compound becomes non-stoichiometric from the original stoichiometric composition. Parameters contributing to the switch of AC order via Au addition are unclear at the moment. What is unequivocal from numerous literatures is that, in the case of Tsai-type materials, the stability of 1/1 and 2/1 ACs as well as QCs are often energetically competitive. Therefore, the incorporation of Au likely alters the free energy landscape in favor of stabilizing the 1/1 AC structure. Assuming Ga, Au and Pt atoms as tri-, mono- and zero-valent, the $e/a$ of the resultant ACs lowers from 1.92 to 1.60 after this transformation. The nominal compositions of the resultant polycrystalline Ga-Pt-Au-Gd 1/1 ACs are listed in Table 1.

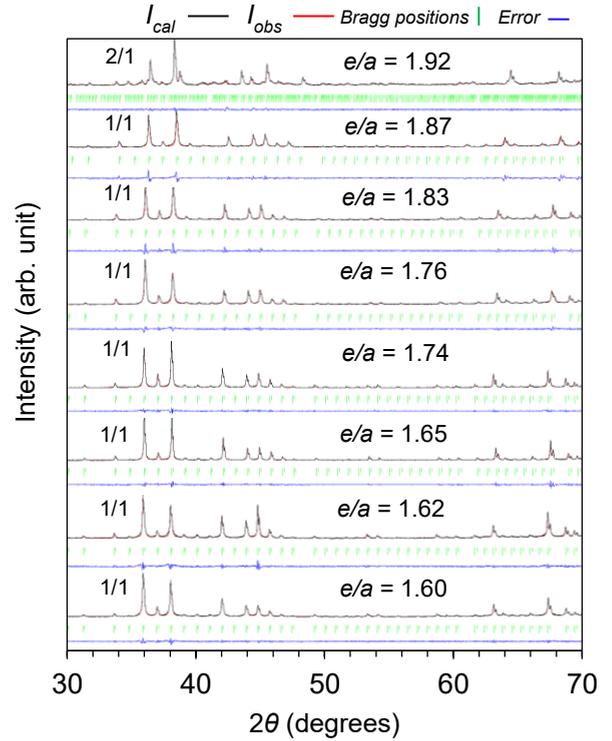

Fig. 1. Le Bail fitting of powder x-ray diffraction (XRD) patterns of the Ga-Pt-Au-Gd ACs annealed at 1073 K. The calculated ($I_{cal}$) and measured ($I_{obs}$) intensities and their difference are represented by red, black and blue, respectively, while the green vertical bars indicate the expected Bragg peak positions.

Figure 1 presents Le Bail fittings of powder XRD patterns of the studied ACs. The fittings were performed using the JANA 2020 software suite [28], assuming space group $Pa\bar{3}$ for the $Ga_{52}Pt_{34}Gd_{14}$ 2/1 AC parent alloy and $Im\bar{3}$ for the quaternary samples. The excellent agreement between the calculated and observed patterns confirms the high structural quality of the samples. It also indicates that the quaternary 1/1 ACs are essentially isostructural in their crystallographic symmetry. Particularly, the structure of the 1/1 ACs accommodates up to 35 at.% Au without affecting the crystallographic symmetry, which is exceptionally large among intermetallic compounds. Monotonous shift of XRD peaks toward low $2\theta$ angles via Au addition indicates lack of preferential sites for Au in the structure.

To confirm assumed symmetries in the Le Bail fittings, a single crystal x-ray diffraction (SCXRD) experiment was carried out on a representative 1/1 AC with the nominal composition $Ga_{33}Au_{33}Pt_{20}Gd_{14}$. Figures S1(a–c) display constructed reciprocal-

Table 1. Nominal compositions, AC order, electron per atom ($e/a$), and magnetic properties of the Ga-Pt-Au-Gd compounds.

| Composition | AC type | $e/a$ | Magnetic state |
|---|---|---|---|
| $Ga_{52}Pt_{34}Gd_{14}$ | 2/1 AC | 1.92 | SG |
| $Ga_{46}Au_{7}Pt_{33}Gd_{14}$ | 1/1AC | 1.87 | AFM |
| $Ga_{43}Au_{12.5}Pt_{30.5}Gd_{14}$ | 1/1AC | 1.84 | SG/FM |
| $Ga_{40}Au_{21}Pt_{25}Gd_{14}$ | 1/1AC | 1.83 | FM |
| $Ga_{35}Au_{29}Pt_{22}Gd_{14}$ | 1/1AC | 1.76 | FM |
| $Ga_{33}Au_{33}Pt_{20}Gd_{14}$ | 1/1AC | 1.74 | FM |
| $Ga_{31}Au_{33}Pt_{20}Gd_{14}$ | 1/1AC | 1.70 | FM |
| $Ga_{30}Au_{33}Pt_{23}Gd_{14}$ | 1/1 AC | 1.65 | FM |
| $Ga_{29}Au_{33}Pt_{24}Gd_{14}$ | 1/1 AC | 1.62 | FM |
| $Ga_{28}Au_{33}Pt_{25}Gd_{14}$ | 1/1 AC | 1.60 | FM |
| $Ga_{25}Au_{35}Pt_{26}Gd_{14}$ | 1/1 AC | 1.52 | SG/FM |

…

space sections perpendicular to the [100], [110], and [111] crystallographic axes, respectively, wherein no violation of the systematic extinction rule for the space group $Im\bar{3}$ was observed confirming the Le Bail fitting results.

Figure S2 in the Supplemental Material shows inverse magnetic susceptibility ($H/M$) of the samples under $\mu_0H = 0.1$ T within a temperature range of 1.8–300K. The results demonstrate a linear behavior of $H/M$ in all samples fitting well to the Curie-Weiss law: $\chi(T) = N_A\mu_{eff}^2\mu_B^2/3k_B(T - \theta_w) + \chi_0$, where $N_A$, $\mu_{eff}$, $\mu_B$, $k_B$, $\theta_w$, and $\chi_0$ denote the Avogadro number, effective magnetic moment, Bohr magneton, Boltzmann constant, Curie-Weiss temperature, and the temperature-independent magnetic susceptibility, respectively. By extrapolating linear least-square fittings within a temperature range of 50–300 K, the $\theta_w$ varies from –10.47 to +14.23 K. Moreover, the $\mu_{eff}$ values are within a range of 7.85 – 8.12 $\mu_B$, close to the calculated value for free $Gd^{3+}$ ions defined as $g_J(J(J+1))^{0.5}$ $\mu_B = 7.94$ $\mu_B$ [29], indicating localization of the magnetic moments on $Gd^{3+}$ ions. The inset of Fig. S2 displays a polynomial fitting of the $\theta_w/dG$ ($dG$ denotes de Gennes parameter) versus $e/a$, wherein a sharp rise below $e/a = 1.9$ followed by a mild fall below $e/a = 1.70$ is clear. This indicates enhanced FM interactions due to Au incorporation into the structure reaching a maximum around $e/a = 1.70$ followed by its subsequent weakening below that.

Figure 2(a) provides temperature dependence of zero-field-cooled (ZFC) dc magnetic susceptibility ($M/H$) of the ACs under

$\mu_0H = 0.01$ T within a temperature range of 1.8 – 20 K. The numbers on top of each curve indicate corresponding $e/a$ values. Clearly, by reducing the $e/a$, the magnetic transition becomes sharper, indicating the onset of spontaneous magnetization. Moreover, in agreement with a trend seen in $\theta_w$, by reducing the $e/a$, the transition temperature rises reaching ~ 15 K followed by its subsequent fall by further reduction of the $e/a$.

Figure 2(b) displays field dependence magnetization ($M$ vs. $H$) of the ACs with varying Au concentrations. Except the spin-glass $Ga_{52}Pt_{34}Gd_{14}$ 2/1 AC, the $M$ of all quaternary 1/1 ACs reaches a full moment of a $Gd^{3+}$ free ion based on Hund's rule (i.e., 7.00 $\mu_B/Gd^{3+}$), though the saturation field ($H_{sat.}$) differs with the $e/a$ of the compound. The inset of Fig. 2(b) provides $e/a$ dependence of $H_{sat.}$. Clearly, by reducing $e/a$, $H_{sat.}$ reaches a minimum of ~ 1 T at $e/a = 1.70$ rising again by further reduction of the $e/a$. A lower $H_{sat.}$ means that the system requires a smaller external magnetic field to fully align the spins, indicating stronger intrinsic magnetic interactions. This aligns with the fact that $\theta_w$ and $T_C$ are also maximized around $e/a = 1.70$. In the present ACs, a sharp rise in magnetic susceptibility and field dependent magnetization indicative of FM order establishment are observed within $e/a$ range of $1.60 - 1.83$ with the borders corresponding to $Ga_{28}Au_{33}Pt_{25}Gd_{14}$ ($e/a = 1.60$) and $Ga_{40}Au_{21}Pt_{25}Gd_{14}$ ($e/a = 1.83$) 1/1 ACs. To confirm the magnetic order establishment in the borders of the FM region, specific heat ($C_p$) measurements are carried out.

Figure 3 displays the temperature-dependence of $C_p$ for the above two samples, evidencing a pronounced anomaly at $T_C = 8.7$ K for both, confirming establishment of FM order within $1.60 \leq e/a \leq 1.83$.

Figure S3 presents the $M^2$ dependence of the $\mu_0(H/M)$ in the form of *standard* Arrott plot [30] for four distinct FM 1/1 ACs with varying $e/a$ ratios of 1.60, 1.71, 1.75, 1.83. The absence of a negative slope and/or an inflection point in Fig. S3 suggests that the transitions are of second-order nature (based on Banerjee criterion [31]). According to the scaling principle, in the second-order phase transition the following relations should hold near $T_C$ [32]:

$$M_s(T) = M_0(-\epsilon)^\beta; \; \epsilon < 0; \; T < T_C \qquad (1)$$

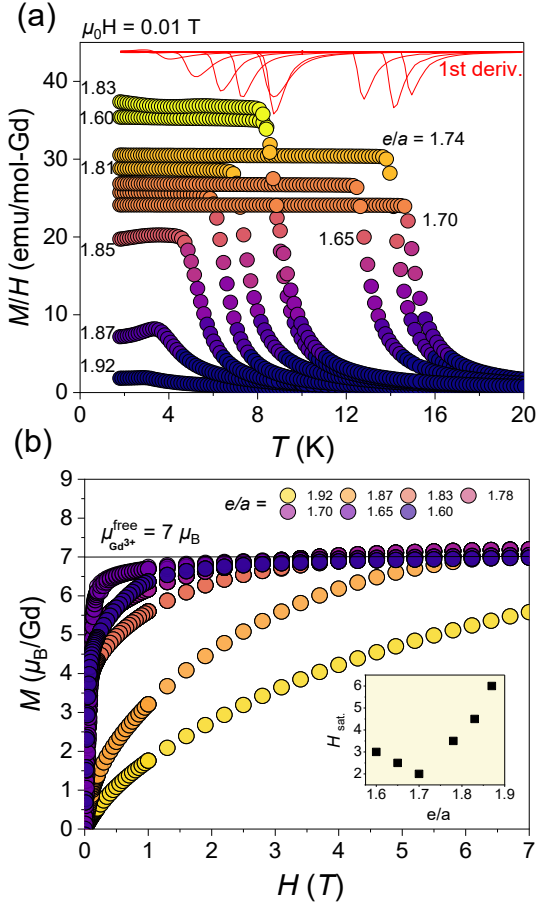

Fig. 2. (a) ZFC magnetic susceptibilities ($M/H$) of the corresponding samples with various e/a values within $1.8 < T < 20$ K measured under $\mu_0\Delta H = 0.01$ T. (b) Field dependence of magnetization for the corresponding samples up to 7 T.

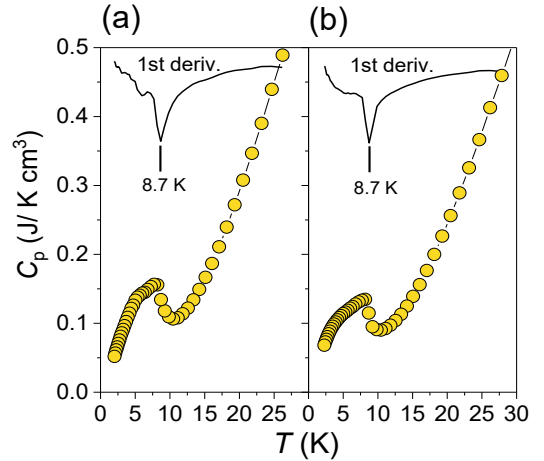

Fig. 3. Temperature dependence of $C_p$ for (a) $Ga_{28}Au_{33}Pt_{25}Gd_{14}$ ($e/a = 1.60$) and (b) $Ga_{40}Au_{21}Pt_{25}Gd_{14}$ ($e/a = 1.83$) 1/1 ACs under 0 T magnetic field.

...

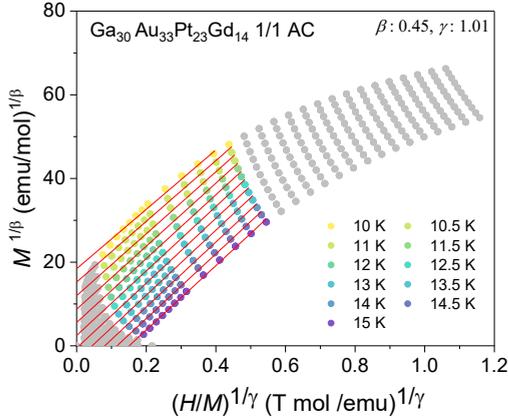

Fig. 4. The modified Arrott isotherms in a form of $M^{1/\beta}$ vs. $H/M^{1/\gamma}$ for Ga$_{30}$Au$_{33}$Pt$_{23}$Gd$_{14}$ 1/1 AC. Nearly parallel linear fittings can be observed within the coloured sections of the isotherms corresponding to magnetic fields of 0.4 T < $H$ < 2.5 T.

$$(H/M)_0(T) = (h_0/M_0)\epsilon^{\gamma}; \ \epsilon > 0; \ T > T_C \qquad (2)$$

where $M_0$ and $h_0$ are critical amplitudes and $\epsilon$ is the reduced temperature $(T - T_C)/T_C$. The critical exponents $\beta$ and $\gamma$ correspond to the spontaneous magnetization $M_s(T)$ below $T_C$ ($H = 0$) and initial inverse magnetic susceptibility $(H/M)_0(T)$ above $T_C$, respectively. Following the approach discussed in the Supplemental Material and Fig. S4 within, the $\beta$ and $\gamma$ within a range of 0.43 – 0.50 and 0.97 – 1.03, respectively, are estimated, as summarized in Table 2. Accordingly, modified Arrott plots are constructed, as shown in Fig. 4 for the Ga$_{30}$Au$_{33}$Pt$_{23}$Gd$_{14}$ 1/1 AC, as an example (Modified Arrott plots for other FM samples are provided in Supplemental Fig. S5). Nearly parallel lines of isotherms in Fig. 4 within a magnetic field range of 0.4 T – 2.5 T (colored sections) with the one close to $T_C$ passing through the origin confirms the credibility of the adopted critical exponents. Using Widom's identity, the estimated $\delta = 1 + \gamma/\beta$ becomes within 2.92–3.39, which are significantly lower than those expected for three-dimensional (3D) universality classes: $\delta = 4.80$ [30] (3D Heisenberg), $\delta = 4.82$ [30] (3D Ising), and $\delta = 5.00$ [31] (tricritical mean-field) but fairly close to the $\gamma/\beta = 3.00$ predicted by the Landau mean-field model [30] (see Table 2) indicating a mean-field nature of the newly developed FM 1/1 ACs near their $T_C$ within a wide $e/a$ space spanning from 1.60 to 1.83.

Next, we investigate $\Delta S_M$ of the present ACs around their transition temperatures. For that purpose, series of temperature dependent field-cooled (FC) magnetization curves within a temperature range of 1.8 – 100 K and magnetic field spanning from 0.01 to 7 T are collected for each sample, as displayed in Fig. S6 of the Supplemental Material. The $\Delta S_M$ is estimated using the thermodynamic Maxwell relation [33]:

$$\Delta S_M(T,H) = \mu_0 \int_{H1}^{H2} \left(\frac{\partial M(T,H)}{\partial T}\right)_H dH \qquad (3)$$

where $M$ and $H$ represent the magnetization and the external magnetic field, respectively. In all ACs, $-\Delta S_M$ exhibits a maximum around transition temperature. The magnitude of the $-\Delta S_M$ peak at each magnetic field is collected from Fig. S6 and summarized in Fig. 5(a), which shows a comprehensive map of $-\Delta S_M$ versus $e/a$. In the map, the left and right vertical axes are $\mu_0 H$ and transition temperature ($T_C$ or $T_f$ depending on the ground state), respectively, with the background colormap representing the magnitude of $|\Delta S_M|$.

Figure 5(a) is informative from several aspects. Notably, it displays two pronounced maxima for the $|\Delta S_M|$ at $e/a \sim 1.60$ and 1.83, corresponding to the boundaries of FM state in the $e/a$ parameter space. At $e/a \sim 1.83$, for instance, the $-\Delta S_M$ reaches $-8.7$ J / K mol-Gd under $\mu_0\Delta H = 5$ T (or 10.35 J/K mol-Gd under $\mu_0\Delta H = 7$ T), marking the highest $|\Delta S_M|$ value ever reported among QCs and ACs. The presence of two maxima in $-\Delta S_M$ at the borders of FM region unveils inverse correlation between $\Delta S_M$ and $T_C$, as commonly seen in R compounds with dominant RKKY interaction [34]. The inverse correlation often follows $\Delta S_M \propto T_C^{-2/3}$ derived from the Weiss mean field equation of state and the Taylor expansion of the Brillouin function ($B_J$) [35] expressed as:

$$-\Delta S_M(T,\mu_0 H) = \frac{1}{2C}\left(\frac{C^2}{K}\right)^{2/3}\left(\frac{\mu_0 H}{T_C}\right)^{2/3} + \cdots , \qquad (4)$$

where $C$ and $K$ are constants depending on the total angular momentum. According to (4), the field dependent magnetic entropy change at $T_C$, i.e., $\Delta S_M(T_C, \mu_0 H)$, scales with $\mu_0 H/T_C$ for the given R element, thus inevitably decreases by increasing $T_C$.

Figure 5(b) plots $-\Delta S_M$ versus $T_C$ (in units of J/ K mol-R) for the present Ga-Au-Pt-Gd 1/1 ACs, along with other ACs and heavy R-based compounds [34] reported elsewhere under $\mu_0\Delta H = 5$ T, which is typically generated by commercial superconducting magnets [36]. The $|\Delta S_M|$ of the present Ga-Au-Pt-Gd 1/1 ACs appear at the far-left side of Fig. 5(b) (represented by red markers) where high $\Delta S_M$ values are typically expected according to equation (4). As shown, the Ga-Au-Pt-Gd 1/1 ACs not only outperform previous ACs by 33% in $\Delta S_M$ magnitude but also showcase comparable $\Delta S_M$ values to other high-performance heavy R-based compounds, such as RCo$_2$ [37] and Er$_5$Si$_4$ [38]. This highlights the potential of the "double hetero-valent elemental substitution" approach used in the present work for tuning magnetic properties and $\Delta S_M$ in stoichiometric compounds, which certainly is a big step forward in material design that could open endless opportunities for designing new high-performance magnetocaloric materials. The important advantage of this approach is that it can be applied to other stoichiometric intermetallic compounds beyond QCs and ACs. In addition, this approach could contribute to reducing synthesis costs by enabling the substitution of constituent elements with more affordable alternatives such as Cu and Ag, which is definitely another important issue in magnetocaloric technology.

Table 2. Comparison of critical exponents ($\beta$, $\gamma$, and $\delta$) and $T_C$ for the present Ga-Pt-Au-Gd FM 1/1 ACs. Theoretical values of the critical exponents for the mean field, tri-critical Mean-Field, 3D Ising and 3D Heisenberg are also provided.

| Composition | $\beta$ | $\gamma$ | $\delta$ | $T_C$ (K) | Reference |
|---|---|---|---|---|---|
| Ga$_{40}$Au$_{21}$Pt$_{25}$Gd$_{14}$ | 0.50 | 0.97 | 2.94 | 8.8 | This work |
| Ga$_{30}$Au$_{33}$Pt$_{23}$Gd$_{14}$ | 0.45 | 1.01 | 3.24 | 12.8 | This work |
| Ga$_{31}$Au$_{35}$Pt$_{20}$Gd$_{14}$ | 0.44 | 0.99 | 3.25 | 14.9 | This work |
| Ga$_{28}$Au$_{33}$Pt$_{25}$Gd$_{14}$ | 0.43 | 1.03 | 3.39 | 8.8 | This work |
| Mean field | 0.50 | 1.00 | 3.00 | – | [30] |
| Tri-critical Mean-Field | 0.25 | 1.00 | 5.00 | – | [31] |
| 3D Ising | 0.325 | 1.24 | 4.82 | – | [30] |
| 3D Heisenberg | 0.365 | 1.386 … | 4.82 | – | [30] |

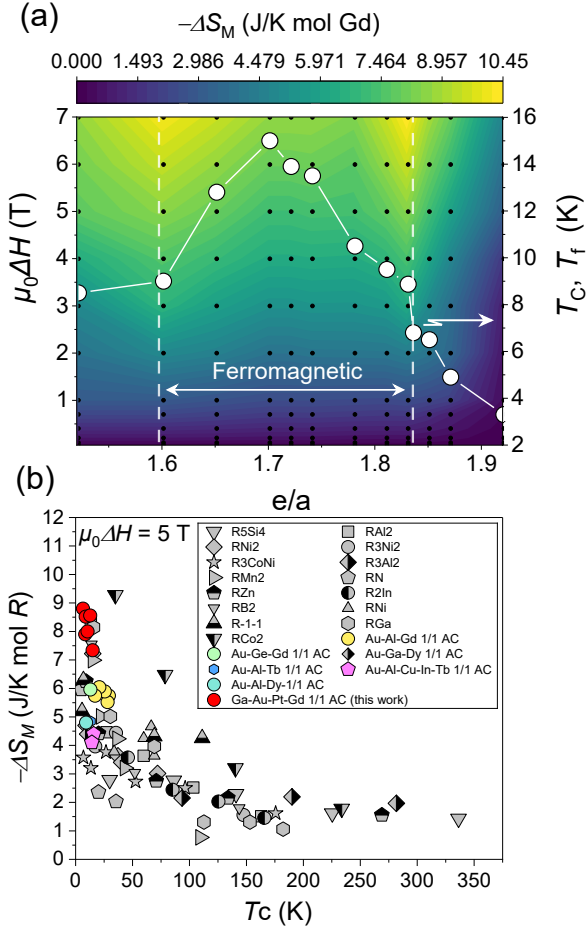

**Fig. 5.** (a) The map of $-\Delta S_M$ versus $\mu_0 H$ and the $e/a$ ratio. The variation of transition temperature $T_C$ estimated from the minimum of the $d(M/H)/dT$ curves is also co-plotted (white dots). (b) The $T_C$ dependance of $-\Delta S_M$ of the present Ga-Pt-Au-Gd 1/1 ACs and several Au-based ACs as well as other heavy R-based compounds under the magnetic field change of 5 T. The data for other R compounds are reproduced from ref. [34]. The unit of $-\Delta S_M$ is J/mol. K (per one mole of R atoms). The shaded part at low temperature region indicates $T_C$ range in Tsai-type compounds.

## VI. CONCLUSION

In summary, we present a novel materials design strategy called a "double hetero-valent elemental substitution" that overcomes the compositional limitation of stoichiometric intermetallic compounds by breaking the structural constraints and expanding the compositional degrees of freedom across a broad valence electron-per-atom ($e/a$) parameter space. By employing this strategy in the stoichiometric Ga–Pt–Gd 2/1 approximant crystals (AC), we demonstrated the emergence of a new family of stable non-stoichiometric 1/1 ACs exhibiting long-range ferromagnetic (FM) order with mean-field-like critical behaviour. Most strikingly, the isothermal magnetic entropy change ($\Delta S_M$) reached a value of $-8.7$ J / K mol-Gd under a 5 T field change at particular $e/a$ values, surpassing all previously reported $\Delta S_M$ values for quasicrystals and ACs. Our findings provide a new material design strategy applicable to any stoichiometric compound with a potential for designing high-performance magnetocaloric materials.

## ACKNOWLEDGMENT

The authors acknowledge Akiko Takeda for assistance in the synthesis of the materials. This work was supported by Japan Society for the Promotion of Science through Grants-in-Aid for Scientific Research (Grants No. JP19H05817, No. JP19H05818, No. JP19H05819, No. JP21H01044, and No. JP24K17016) and Japan Science and Technology agency, CREST, Japan, through a grant No. JPMJCR22O3.